\documentstyle[pre,aps]{revtex}

\begin{document}
\draft
\author{S. Gluzman$^1$ and V. I. Yukalov$^2$\footnote{the author to whom the
correspondence to be addressed}}
\address{$^1$International Center of Condensed Matter Physics\\
University of Brasilia, CP 04513, Brasilia, DF 70919-970, Brazil \\
and \\
$^2$Bogolubov Laboratory of Theoretical Physics \\
Joint Institute for Nuclear Research, Dubna 141980, Russia}
\title{Resummation Methods for Analyzing Time Series}

\maketitle

\begin{abstract}
An approach is suggested for analyzing time series by means of resummation
techniques of theoretical physics. A particular form of such an analysis,
based on the algebraic self-similar renormalization, is developed and
illustrated by several examples from the stock market time series.
\end{abstract}

\vspace{3cm}


\section{Self-Similar Renormalization}

Many data in different sciences are presented in the form of time series.
The problem of analyzing the latter consists in understanding the dynamics
of motion from one temporal point to another in the past and, hopefully, in
forecasting the data for at least some near future. The standard approach to
time-series analysis is to try to guess what stochastic dynamical system is
behind the data, that is, to attempt to model the events producing the
considered data by stochastic differential or finite-difference equations
[1,2]. Such an approach has been proved to be reasonable when applied to a
system close to a stationary state, but sudden changes in the behavior of
the dynamical system generating the time series cannot be accurately grasped.

In this communication we advance a novel approach to analyzing time 
series, based on resummation techniques that are used in theoretical
physics. As examples for illustration, we opt for time series generated by
economic systems, including sharp changes in their behavior. Let us note 
that different physical analogies and techniques are now often used for 
describing ecomonic phenomena [3-8].

As a starting point, we present a discrete set of data, available from the
system past, in the form of a polynomial ( i.e. as a formal power series),
which is a direct analog of a perturbative expansion valid as time 
$t\rightarrow 0$. The rest consists in the application of one or another
resummation technique to the asymptotic expansion where the role of a
coupling constant is played by time. If only a few points from the system
history are known, the degree of the asymptotic expansion will be low, and
the most popular Pade$^{\prime }$ summation [9] is very difficult, if
possible, to apply. However, there is an approach called the 
self--similar approximation theory [10-17] which successfully works even 
for a small number of asymptotic terms. Here we shall use a variant of 
this approach, the algebraic self--similar renormalization [15-17]. We 
consider below several examples taken from the history of real stock markets 
using only a few, up to six, points from the history in order to calculate 
the following values of the share prices.

First, let us give the general scheme of the method we suggest. Assume that
the values of the sought function, $f(t)$, are known for $n$ equidistant
successive moments of time, $t=k=0,1...,n-1$, so that 
\begin{equation}
f(0)=a_{0},\ \ f(1)=a_{1},\ ...,\ f(n-1)=a_{n-1}.  \label{1}
\end{equation}
Let us be interested in the value of the function $f(n)$ in the following 
$n$--th moment of time. To proceed further, it is important to find a 
compact representation of the set of the data (1) in the form of an 
explicit function. To this end, we can always use a formal Taylor series 
of the sought function, $f(t)=\sum_{k=0}^{\infty }A_{k}t^{k}$. Since only a 
finite set of values (1) is available, one can reconstruct only a finite set 
of coefficients, $A_{k}=A_{k}(a_0,a_1,\ldots,a_{n-1}),\; k=0,1,\ldots,n-1$, 
from condition (1). Thus, we obtain an approximate expansion 
\begin{equation}
f(t)\simeq \sum_{k=0}^{n-1}A_{k}(a_0,a_1,\ldots,a_{n-1})\ t^k .  
\label{2}
\end{equation}
Let us stress that expansion (2) has no direct sense if continued 
straightforwardly to the region of finite arbitrary $t$. The
problem of reconstructing the value of a function in some distant moment of
time from the knowledge of a finite set of its values in preceding time
moments becomes now equivalent to the reconstruction of the function for the
finite value of its argument, $t=n$, from the knowledge of its asymptotic
expansion as $t\rightarrow 0$. In theoretical physics, such a problem is
called renormalization or resummation problem [9,18]. Thus, the problem of
forecasting the future values from the set of historical data given in the
form of time series becomes equivalent to the renormalization 
(resummation) of asymptotic series. An analytical tool for the solution 
of this problem, called the algebraic self-similar renormalization, has 
been recently developed [15-17]. We describe here only its principal points 
that are important within the context of this letter. The polynomial 
representation (2) gives for the sought function $f(t)$ the following 
$n$ approximations $p_{i}(t),\; i=0,1,...,n-1$,
\begin{equation}
p_{0}(t)=A_{0}\equiv a_{0},\ \ p_{1}(t)=p_{0}(t)+A_{1}t,...,\ \
p_{n-1}(t)=p_{n-2}(t)+A_{n-1}t^{n-1}.  \label{3}
\end{equation}
The algebraic self--similar renormalization starts from applying to the
approximations (3) an algebraic transformation, defining 
$P_{i}(t,s)=t^{s}p_{i}(t),\ i=0,1,...,n-1$, with $s\geq 0$. This
transformation raises the powers of series (3), and allows us to take
effectively into consideration more points from the system history. We use
in what follows the strongest form of such a transformation when 
$s\rightarrow \infty$, which results in a nice exponential
representation for the sought function. The sequence of so transformed 
approximations $P_{i}(t,s)$ is considered as a dynamical system in 
discrete time $i=0,1,...,n-1$. In order to describe the system evolution 
with time, we introduce, according to Refs. [10-14], a new variable 
$\varphi $ and define the so-called  expansion function $t(\varphi ,s)$ 
from the equation $P_{0}(t,s)=a_{0}t^{s}=\varphi$, which results in 
$t(\varphi ,s)=\left( \varphi /a_{0}\right) ^{1/s}$. Then we construct an 
approximation cascade [10-14] whose trajectory points are given by the
expressions $y_{i}(\varphi ,s)\equiv P_{i}(t(\varphi ,s),s)$. 
Embedding this cascade into an approximation flow, one can write the 
evolution equation in the form of the functional self--similarity 
relation $y_{i+p}(\varphi,s)=y_{i}(y_{p}(\varphi ,s),s)$. At 
this stage, we can check the effectiveness of the algebraic 
transformation by calculating the local multipliers,
\begin{equation}
m_{i}(t,s)\equiv \left [
\frac{\partial y_{i}(\varphi ,s)}{\partial\varphi }
\right ]_{\varphi = P_{0}(t,s)} \; , 
\label{4}
\end{equation}
as $s\rightarrow \infty $. For calculations we use the integral form of 
the self-similarity relation,
$$ 
\int_{P_{i-1}}^{P_{i}^{*}}\frac{d\varphi }{v_{i}(\varphi ,s)}=\tau ,
$$
where the cascade velocity $v_{i}(\varphi,s)=y_{i}(\varphi,s)-
y_{i-1}(\varphi ,s)$ and $\tau$ is the minimal number of steps of 
the procedure needed to reach the fixed point $P_{i}^{*}(t,s)$ of the 
approximation cascade. It is possible to find $P_{i}^{*}(t,s)$
explicitly and to perform an inverse algebraic transform after which the
limit $s\rightarrow \infty $ is to be taken. The first step of the
self--similar renormalization is completed. Then the procedure can be 
repeated as many times as it is necessary to renormalize all polynomials. 
This is the main idea of the self-similar bootstrap [17]. Accomplishing this 
program, we come to the following sequence of the self-similar exponential 
approximants 
\begin{equation}
f_{j}^{*}(t,\tau )=A_{0}\exp \left( \frac{A_{1}}{A_{0}}t\exp \left( \frac{
A_{2}}{A_{1}}t...\exp \left( \frac{A_j}{A_{j-1}}\tau t\right) \right)
...\right) ,\ \ j=1,2,...,n-1.  \label{5}
\end{equation}
In order to check whether this sequence converges, we have to analyze the
corresponding mapping multipliers. From the equation $f_{1}^{*}(t,1)=\varphi$,
we find $t(\varphi)=(A_{0}/A_{1})\ln(\varphi /A_{0})$. Then we construct an 
approximation cascade, as is described above, getting $z_j(\varphi,\tau)
\equiv f_{j}^{*}(t(\varphi),\tau)$, and define 
\begin{equation}
M_{j}(t,\tau )\equiv \left [
\frac{\partial z_{j}(\varphi ,\tau )}{\partial \varphi }
\right ]_{\varphi =f_{1}^{*}(t,1)}.  
\label{6}
\end{equation}
Selecting from the sequence of $f_{j}^{*}$ two successive terms with
smallest $M_{j}(t)\equiv M_{j}(t,1)$, as a rule the two last terms of the
sequence, we can finally determine $\tau \ $ from the minimal velocity
condition written as the minimal difference condition 
\begin{equation}
\min_{\tau }\left| f_{j}^{*}(t,\tau )-f_{j-1}^{*}(t,\tau )\right| ,
\label{7}
\end{equation}
whose solution $\tau =\tau (t)$ allows us to write the final self-similar
exponential approximation for the sought function
\begin{equation}
f_{j}^{*}(t)=f_{j}^{*}(t,\tau (t)).
\end{equation}

The approximants $f_{n-2}^{*}(n,1)$ and $f_{n-1}^{*}(n,1)$ usually frame the
optimal value $f_{n-1}^{*}(n)$. One should analyze, following the procedure
described above, a set of self-similar approximations for different 
number of points from the history, i.e., one has to calculate several 
possible $f_{j}^{*}(t),\ j=2,3,\ldots,n-1$, and to choose among them 
that which corresponds to the smallest mapping multiplier calculated at 
the fixed point. The value of the multiplier at the fixed point can be  
defined as 
\begin{equation}
M_{j}^{*}(t)\equiv \frac{M_{j}(t,\tau (t))+M_{j-1}(t,\tau (t))}{2}.
\end{equation}

Now we pass to the illustration of the method by examples from the 
history of various stock markets. We concentrate on different stock 
market crises, when the prices changed sharply during the period of time 
comparable to the resolution of time series. The data, unless stated  
otherwise, are taken from the books of International Financial
Statistics issued by the International Monetary Fund.

\section{Mechanism of Crash}

Consider the behavior of the Dow Jones index in the vicinity of the 
crisis of October 27, 1997. We intend to give an illustration of the  
self-similar renormalization scheme presented above and also to make 
some general remarks concerning the mechanism of crash. We are going to  
make a forecast for the Dow Jones index for {\it October 25, 1997} based 
on different number of points from the system history. 

{\bf Three point forecast}. The following values are available in the 
period of time from ${\sl September\ 13,1997}$
till ${\sl October\ 11,1997}$ taken with the two-week resolution (data are
taken from the Yahoo Finance chart): 
$$
a_{0}=7720\ (Sept.\ 13),~\quad a_{1}=7920,\quad a_{3}=8000\ (Oct.\ 11).
$$
The coefficients of the polynomial (2) are $A_{1}=260$ and $A_{2}=-60$. 
We find the exponential approximants $f_{1}^{*}(3,1)=8540.8$ and 
$f_{2}^{*}(3,1)=8120.4$ and the mapping multipliers $\ M_{1}(3)=1$ and 
$M_{2}(3)=0.146$. From the minimal difference condition $\min_{\tau }\left|
f_{2}^{*}(3,\tau )-f_{1}^{*}(3,\tau )\right| $,\ we obtain $\tau =0.641428$.
Finally, the self-similar exponential approximation is $f_{2}^{*}(3)=8237$ 
and the modulus of the multiplier (9) is 
$ \left| M_{2}^{*}(3)\right| =0.485.$

{\bf Four--point forecast}. The following values are available in the period 
of time from ${\sl August\; 30,\; 1997}$ till ${\sl October\; 11,\; 1997}$ 
taken with the two week resolution (data are taken from the Yahoo Finance 
chart): 
$$
a_{0}=7690\; (Aug.\; 30), \quad a_{1}=7720, \quad a_{2}=7920,\quad 
a_{3}=8000\ (Oct.\; 11).
$$
The coefficients of the polynomial are $A_{1}=-151.667,\; A_{2}=230$,  
and $A_{3}=-48.333$. Repeating the same procedure as above, we find the 
exponential approximants $f_{2}^{*}(4,1)=7689$ and $f_{3}^{*}(4,1)=7646$, 
with the mapping multipliers $\ M_{2}(4)=-0.013$ and $M_{3}(4)=0.046$. From 
the minimal difference condition  
$\min_{\tau }\left|f_{3}^{*}(4,\tau )-f_{2}^{*}(4,\tau )\right| $, we obtain 
$\tau =0.6026$. The self--similar exponential approximant becomes 
$f_{3}^{*}(4)=7674$ and the corresponding multiplier modulus is $\left| 
M_{3}^{*}(4)\right| =0.048$.

{\bf Five--point forecast}. Consider the historical data for the Dow Jones
index in the period of time from ${\sl August\; 16,\; 1997}$ to 
${\sl Octobert\; 11,\; 1997}$. The following data are available: 
$$
a_{0}=7780\ ({\sl Aug.\ 16\ }),\quad a_{1}=7690,\quad a_{2}=7720,\quad
a_{3}=7920,\quad a_{4}=8000\ ({\sl Oct.\ 11}).
$$
From condition (1), the coefficients of the polynomial (2), with 
$i=0,1,...,4$, are 
$$
A_{1}=-48.333,\quad A_{2}=-120.833,\quad A_{3}=93.333,\quad A_{4}=-14.167.
$$
The following three higher-order self-similar exponential approximants, 
$f_{2}^{*}(t,\tau )$, $f_{3}^{*}(t,\tau )$, and $f_{4}^{*}(t,\tau)$ 
can be written, so that at $t=5$ we have $f_{2}^{*}(5,1)=0$, 
$f_{3}^{*}(5,1)=7472$, and $f_{4}^{*}(5,1)=6113$. The corresponding 
mapping multipliers, $M_{2}(5)=0,\; M_{3}(5)=$ $0.319$, and 
$M_{4}(5)=13.569$, signal that the subsequence $f_{2}^{*},\ f_{3}^{*},\
f_{4}^{*}$ is unstable. But from the minimal difference condition 
$\min_{\tau }\left| f_{4}^{*}(5,\tau )-f_{3}^{*}(5,\tau )\right| $ we can
still locate the fixed point and find $\tau =0.6232$. The self--similar
exponential approximation is $f_{4}^{*}(5)=7069$, with the multiplier
modulus  $\left| M_{4}^{*}(5)\right| =0.163$.

{\bf Six--point forecast}. Consider the data for the Dow Jones index in 
the period of time from ${\sl August\; 2,\; 1997}$ to ${\sl October\; 11, 
\; 1997}$: 
$$
a_{0} = 8170\ (Aug.\ 2), \quad  a_{1}=7780, \quad a_{2}=7690, $$
$$
a_{3}=7720, \quad a_{4}=7920, \quad a_{5} =8000 \quad ({\sl Oct.\ 11}).
$$
The coefficients of the polynomial are $A_{1}=-771.5,\;A_{2}=582.917,\; 
A_{3}=-253.75,\; A_{4}=57.083$, and $A_{5}=-4.75$. Following the standard 
prescriptions, we can find the exponential approximants 
$f_{4}^{*}(6,1)=7720$ and $f_{5}^{*}(6,1)=7138$ and the mapping
multipliers $M_{4}(6)=-0.308$ and $M_{5}(6)=0.198$. From the minimal
difference condition $\tau =\exp\left( 6A_{5}\tau \ /A_{4}\right) $ we
obtain $\tau =0.7037$. For the self--similar exponential approximation we 
get $f_{5}^{*}(6)=7328$, with the multiplier modulus  
$\left|M_{5}^{*}(6)\right| =0.087$.

All forecasts, except the three-point one, predict a decrease of the Dow
Jones index. Normally, in the absence of a dramatic event which can alter  
drastically the market activity, one should expect that the most stable 
trajectory is to be realized. Expecting the sequence of the multipliers 
$\{M_{j}^{*}\},\; j=2,3,..,5$, we notice that the four--point forecast 
$f_{3}^{*}(4)=7674$ is optimal from the viewpoint of stability, that is, 
it corresponds to the minimal mapping multiplier, and, thus, this 
four-point prediction is to be chosen as the final estimate. The Dow 
Jones index on {\it October 25, 1997} was $7715$. The percentage error of 
our forecast equals\ $-0.531\%$. An upward development, predicted by the 
three--point forecast, appears to be less stable than the decaying 
trajectories. The level achieved by the market after the bubble burst 
corresponds rather to a correction than to a crash.

When the normal evolution is disrupted by some unexpected {\it negative}
dramatic events, such as the crisis in Hong Kong, which disturbs the
self-similar dynamics, then it may happen that the market temporarily 
selects not the most stable trajectory. In its search for a solution, the 
market may leave for a while the most stable trajectory and 
jump to a less stable one. Then, it can bounce back to the most stable  
trajectory or it can fall for a while on the least stable trajectory 
corresponding, in this case, to the five--point forecast. During a short 
period of time, less than the time--series resolution, one can observe a 
rapid succession of all possible trajectories. After the source of
the self--similarity violation is lifted, one may expect that the market 
will return to the most stable trajectory. The motion to the least 
stable trajectory is the most dramatic moment, as is seen from the  
analysis of the five--point forecast. Since both starting polynomial 
coefficients are negative and $\left|A_{2}\right| >>\left| A_{1}\right|$, 
an extremely strong tendency to decay, a sort of panic, appears 
resulting in $f_{2}^{*}(5,1)=0$; however the situation is remedied due to  
the higher--order coefficient $A_{3}>0$, which reflects still existing 
moderately optimistic views; though such an optimism remains limited by 
pessimistic views represented by the negative $A_{4}$. The finite value  
given by $f_{4}^{*}(5)=7069$ is quite low because of such a combination 
of tendencies. We would like to stress that in the case of the five-point 
forecast the fixed point determined by the locally unstable
sequence $f_{3}^{*},\; f_{4}^{*}$ lies very close to the boundary
separating stable from unstable trajectories. This is why the short-term  
market changes can be vary rapid, first, to the level determined by 
$f_{5}^{*}(6)=7328$, and then to the level $f_{3}^{*}(4)=7674$. We conclude 
that the following reasons caused the October 1997 fluctuations: (1) the  
burst of the upward bubble accompanied by the appearance of several 
stable decaying trajectories; (2) the location of one of these 
trajectories, corresponding to strong decay, at the boundary between 
stable and unstable trajectories; (3) the Hong Kong crash that pushed  
the market from the most stable trajectory with moderate corrections
right to this unstable trajectory, although shortly after this crisis  
the market promptly bounced back to the nearby stable trajectories.

\section{Market Crises}

We shall consider several examples of crises happened in different stock  
markets. In all the cases we have performed the same steps as in Sections 
I and II, but only the optimal forecasts are presented below. It turns 
out that different crises require for their description different number 
of historical points. Let us start from the crises which can be described 
knowing only {\bf three points} from the system history:
 
$(i)$ Consider the dynamics of the average index of the French industrial 
share prices from ${\sl 1986 }$ to ${\sl 1988}$, with one year time 
resolution, and make a forecast for ${\sl 1989}$. The following 
historical data are available (the price in ${\sl 1985}$ is taken for  
$100,{\sl 1985=100}$) : 
$$
a_{0}=153.3\ ({\sl 1986}),\quad a_{1}=177.6\ ({\sl 1987}),\quad a_{2}=162.1\
({\sl 1988}). 
$$
From condition (1), the coefficients of polynomial (2), with $i=0,1,2$,  
can be found giving $A_{1}=44.2$ and $A_{2}=-19.9$. The following
exponential approximants (5) can be readily obtained: 
$$
f_{1}^{*}(t,\tau )=a_{0}\exp \left( \frac{A_{1}}{a_{0}}\tau t\right) ,
\qquad
f_{2}^{*}(t,\tau )=a_{0}\exp \left( \frac{A_{1}}{a_{0}}t\exp \left( \frac{%
A_{2}}{A_{1}}\tau t\right) \right) ,  
$$
from where, at $t=3$ and $\tau =1$, we have $f_{1}^{*}(3,1)=364.078$ 
and  $f_{2}^{*}(3,1)=191.8$. The mapping multipliers can be 
calculated using formula (6), which yields $M_{1}(3)\equiv 1$ and 
$M_{2}(3)=-0.048$, so that the sequence  $f_{1}^{*},\; f_{2}^{*}$ is locally
stable [14]. From the minimal difference condition $\min_{\tau }\left|
f_{2}^{*}(3,\tau )-f_{1}^{*}(3,\tau )\right| $, which reads as 
$$\tau =\exp \left( \frac{3A_{2}}{A_{1}}\tau \right) , $$
we find $\tau =0.505$. Finally, the self-similar exponential 
approximation for the sought share price index is $f_{2}^{*}(3)=237.34$. 
The actual value of the index in ${\sl 1989}$ was $234.9$. The 
percentage error of our forecast is $1.039\%$.

$(ii)$ In $1{\sl 990-1992},$ the average index of the Denmark shipping share
prices had the following values (${\sl 1990}=100$) : 
$$
a_{0}=100\ (19{\sl 90}),\quad a_{1}=100\ (19{\sl 91}),\quad a_{2}=92\ ({\sl %
1992}). 
$$
Let us look for the price in ${\sl 1993}$. This case can be considered by
analogy with the previous one. The polynomial coefficients can be readily
found being $A_{1}=4$ and $A_{2}=-4$. For the considered index we 
have $f_{1}^{*}(3,1)=112.75$ and $f_{2}^{*}(3,1)=100.60$. The
mapping multiplier $M_{2}(3)=-0.089$, hence the sequence of exponential
approximants locally converges. From the minimal difference condition,
we find $\tau =0.35$. Thus, the self-similar exponential approximation 
is  $f_{2}^{*}(3)=104.289.$ The actual index in ${\sl 1993}$ was $105$, 
so the error is $-0.677\%$.

$(iii)$ Consider the average index of the Swiss industrial share prices from 
${\sl 1993}\ $to ${\sl 1995}$ (${\sl 1990}=100$): 
$$
a_{0}=137.4\ ({\sl 93}),\quad a_{1}=159.2\ ({\sl 94}),\quad a_{2}=166\ ({\sl %
95}). 
$$
The corresponding polynomial coefficients are $A_{1}=29.3$ and 
$A_{2}=-7.5$. Let us make a forecast for ${\sl 1996}$. Repeating literally 
the same steps as above, we can easily calculate the exponential 
approximants  $f_{1}^{*}(3,1)=260.508,\; f_{2}^{*}(3,1)=184.883$, and 
the mapping multiplier $M_{2}(3)=0.076$. Then from the minimal 
difference condition we find $\tau=0.621$. So, the self-similar 
exponential approximation is $f_{2}^{*}(3)=204.394$. The price in 
${\sl 1996}$ was $206.3$, which shows that the percentage error of our 
forecast equals\ $-0.924\%$.

We continue with the crises which can be described using {\bf four--points}
from the history:

$(i)$ Consider the dynamics of the average index of the French industrial 
share prices in ${\sl 1970-1973}$, with one year resolution $\ $(${\sl 
1970}=100$): 
$$
a_{0}=100\ ({\sl 70}),\quad a_{1}=95.8\ ({\sl 71}),\quad a_{2}=107.4\ ({\sl %
72}),\quad a_{3}=129.7\ ({\sl 73}). 
$$
Let us make a forecast for ${\sl 1974}$. The polynomial coefficients found 
from condition (2) are $A_{1}=-13.8,\; A_{2}=10.45$ and $A_{3}=-0.85$. 
The exponential approximants (5) are 
$$ 
f_{1}^{*}(t,\tau )=a_{0}\exp \left( \frac{A_{1}}{a_{0}}\tau t\right) ,
$$
$$
f_{2}^{*}(t,\tau )=a_{0}\exp \left( \frac{A_{1}}{a_{0}}t\exp \left( \frac{%
A_{2}}{A_{1}}\tau t\right) \right) ,  
$$
$$
f_{3}^{*}(t,\tau )=a_{0}\exp \left( \frac{A_{1}}{a_{0}}t\exp \left( \frac{
A_{2}}{A_{1}}t\exp \left( \frac{A_{3}}{A_{2}}\tau t\right) \right) \right) .
$$
Therefore, at $t=4$, we have $f_{1}^{*}(4,1)=57.58,\; f_{2}^{*}(4,1)=97.366$, 
and $f_{3}^{*}(4,1)=93.996$. The corresponding multipliers are 
$M_{1}(4)\equiv1,\quad M_{2}(4)=-0.166$, and $M_{3}(4)=-0.087$. From 
the minimal difference condition $\min_{\tau }\left| f_{3}^{*}(4,\tau
)-f_{2}^{*}(4,\tau )\right|$, which reads 
$$ 
\tau =\exp \left( \frac{4A_{3}}{A_{2}}\tau \right) , 
$$
we find $\tau =0.7767$.$\ $Thus, the self-similar exponential approximation
is $f_{3}^{*}(4)=94.885.$ Our estimate agrees well with the actual price
equal to $96.6$ in ${\sl 1973}$. The percentage error is $-1.775\%$.

$(ii)$ In ${\sl 1985-1988},$ the average index of the Netherlands general
share prices had the following values: 
$$
a_{0}=100\ ({\sl 85}),\quad a_{1}=128.7\ ({\sl 86}),\quad a_{2}=129.2\ ({\sl 
87}),\quad a_{3}=119.7\ ({\sl 88}). 
$$
We are going to forecast the price for ${\sl 1989}$. The polynomial 
coefficients are $A_{1}=48.867,\; A_{2}=-23.2$, and $A_{3}=3.033$. The local 
multiplier $m_{1}(t,s)=1+A_{1}\left( 1+s\right)t /A_{0}s$, as 
$s\rightarrow \infty$ and $t=4,$ takes the value 
$m_{1}(4,\infty )=1+4A_{1}/A_{0}=2.995$, hence the quality of the sequence of 
$P_{i},\ i=0,1,...,3$, is not good. In this case, we can
consider another sequence of approximations, which does not include the
constant term. The expansion function now is determined from the
equation $A_{1}t^{1+s}=\varphi$ giving $t(\varphi ,s)=\left( \varphi
/A_{1}\right) ^{1/\left( 1+s\right) }$. The corresponding local multiplier,
defined by the formula analogous to (4), that is $m_{3,1}(t,s)=1+A_{2}\left(
2+s\right) \; t /A_{1}\left( 1+s\right)$, as $s\rightarrow \infty ,\;
t=4$, equals $-0.899<1$. Thence the following self-similar exponential
approximants can be obtained (see Ref. [17] for more details): 
$$
f_{3}^{*}(t,\tau )=a_{0}+A_{1}t\exp \left( \frac{A_{2}}{A_{1}}\tau t\right) ,
$$
$$
f_{4}^{*}(t,\tau )=a_{0}+A_{1}t\exp \left( \frac{A_{2}}{A_{1}}t\exp \left( 
\frac{A_{3}}{A_{2}}\tau t\right) \right) .  
$$
This leads to the estimates $f_{3}^{*}(4,1)=129.264$ and 
$f_{4}^{*}(4,1)=163.417$. The mapping multiplier $M_{3}(4)$, calculated from
the formula analogous to (6), equals $0.613$, which tells that the sequence 
$f_{3}^{*}$, $f_{4}^{*}$ is stable. From the minimal difference 
condition $\min_{\tau }\left| f_{4}^{*}(4,\tau )-f_{3}^{*}(4,\tau )\right|$   
we find $\tau =0.6952$, which results in the self-similar exponential
approximation $f_{4}^{*}(4)$ $=152.206$. Our forecast agrees well with the
index in ${\sl 1989}$, equal to $151.4$, thus the error being $0.532\%$.

$(iii)$ Consider the data for the average index of the Mexican share prices
in $\ {\sl 1991-1994}$ $({\sl 1990}=100):$
$$
a_{0}=190.1\ ({\sl 91}),\quad a_{1}=291.3\ ({\sl 92}),\quad a_{2}=325.6\ 
( {\sl 93}),\quad a_{3}=442.1\ ({\sl 94}). 
$$
The polynomial coefficients are $A_{1}=184.35,\; A_{2}=-108$, and 
$A_{3}=24.85$. Let us make a forecast for ${\sl 1995}$. Since $m_{1}(4,\infty
)=4.879$ and $m_{3,1}(4,\infty )=-1.343$, we will proceed in a close 
analogy to the case ($ii$). Repeating the same steps as above, we can readily
calculate the exponential approximants $f_{3}^{*}(4,1)=260.893$ and 
$f_{4}^{*}(4,1)=480.018$ and the mapping multiplier $M_{3}(4)=0.13$. Then
from the minimal difference condition, we get $\tau =0.584$. The 
self--similar exponential approximation is $f_{4}^{*}(4)=377.695$. The 
index in ${\sl 1995}$ was $389.3$; so the percentage error of our forecast  
equals $-2.981\%$.

{\bf Five--point} crises: $(i)$ Consider the historical data for 
the average index of the German industrial share prices in ${\sl 1957-1961}$,
with one year resolution, and let us make a forecast for ${\sl 1962}$.  
The following data are available (${\sl 1958}=100$) : 
$$
a_{0}=78\ ({\sl 57}),\quad a_{1}=100\ ({\sl 58}),\quad a_{2}=171\ ({\sl 59}%
),\quad a_{3}=272\ ({\sl 1960}),\quad a_{4}=282\ ({\sl 61}). 
$$
From condition (1), the coefficients of polynomial (2), with 
$i=0,1,...,4$, are $A_{1}=16.667,\; A_{2}=-12.75,\;
A_{3}=22.333$, and $A_{4}=-4.25$. The higher-order self-similar
exponential approximants are
$$
f_{3}^{*}(t,\tau )=a_{0}\exp \left( \frac{A_{1}}{a_{0}}t\exp \left( \frac{%
A_{2}}{A_{1}}t\exp \left( \frac{A_{3}}{A_{2}}\tau t\right) \right) \right) ,
$$
$$
f_{4}^{*}(t,\tau )=a_{0}\exp \left( \frac{A_{1}}{a_{0}}t\exp \left( \frac{%
A_{2}}{A_{1}}t\exp \left( \frac{A_{3}}{A_{2}}t\exp \left( \frac{A_{4}}{A_{3}}%
\tau t\right) \right) \right) \right) . 
$$
At $t=5$, we get $f_{3}^{*}(5,1)=226.885$ and $f_{4}^{*}(5,1)=199.314$.  
The corresponding mapping multipliers, $M_{3}(5)=1.003$ and 
$M_{4}(5)=0.687$, signal that the subsequence $f_{3}^{*},\; f_{4}^{*}$ is 
stable. From the minimal difference condition $\min_{\tau }\left| 
f_{4}^{*}(5,\tau )-f_{3}^{*}(5,\tau )\right| $  we have 
$$ \tau =\exp \left( \frac{5A_{4}}{A_{3}}\tau \right) .  $$
The latter equation gives $\tau =0.577$. The self--similar exponential 
approximation $f_{4}^{*}(5)=221.268$ agrees well with the actual value 
of the index in ${\sl 1962}$ equal to $221$. The percentage error is 
$0.121\%$.

$(ii)$ The average index of the Swedish share prices in the period of time
from {\sl May $1987$} to {\sl September $1987$} had the following 
values (${\sl 1980}=100$): 
$$
a_{0}=726\ ,\quad a_{1}=740\ ,\quad a_{2}=801\ ,\quad a_{3}=825,\quad
a_{4}=877\ . 
$$
What was the value of the index in {\it October 1987} ? This case can be
considered by analogy with $(i)$. The polynomial coefficients are 
$A_{1}=-74.75,\; A_{2}=133.792,\; A_{3}=-51.25$, and 
$A_{4}=6.208$. The estimates for the index are $f_{3}^{*}(5,1)=632.564$
and $\ f_{4}^{*}(5,1)=710.11.$ The multipliers are $M_{3}(5)=0.861$ and
$M_{4}(5)=-0.06$, hence we conclude that the sequence of the exponential 
approximants locally converges. From the minimal difference condition, 
we get $\tau =0.6675$. The self--similar exponential approximation 
$f_{4}^{*}(5)=695.732$. The actual value of the index in {\it October 1987}
was $697$. So that the error of our prediction is $-0.182\%$.

$(iii)$ Consider the historical data for the average index of the Israel
industrial share prices in ${\sl 1989-1993}$ (${\sl 1990}=100$): 
\[
a_{0}=87\ ({\sl 89}),\ \ a_{1}=100\ ({\sl 90}),\ \ a_{2}=155\ ({\sl 91}),\ \
a_{3}=297\ ({\sl 92}),\ \ a_{4}=418\ {\sl (93}). 
\]
The polynomial coefficients are $A_{1}=45.25,\; A_{2}=-71.625,\;
A_{3}=45.75$, and $A_{4}=-6.375$. Let us make a forecast for the year 
${\sl 1994}$. Repeating the same steps as above, we find the 
approximants  $f_{3}^{*}(5,1)=569.963$ and $f_{4}^{*}(5,1)=146.153$ and 
the mapping multipliers $M_{3}(5)=0.608$ and $M_{4}(5)=0.004$. Then
from the minimal difference condition, we get $\tau =0.64$. The self-similar
exponential approximation is $f_{4}^{*}(5)=221.18$. The actual value of the
index in {\sl $1994$ }was $257$. The percentage error of our forecast
equals $-13.938\%$.

{\bf Six--point }crises: $(i)$ Consider the historical data for the
Standard Statistics index of the New York Stock Exchange prices in the
period from {\sl April $1929$} till {\sl September $1929$}, with one month
resolution, and let us make a forecast for {\it October }${\sl 1929}$.
The following historical data are available (taken from the League of
Nations Statistical Yearbook; ${\sl 1926}=100$) : 
$$
a_{0}=193\ (Apr.\; {\sl 1929}),\ \ a_{1}=193,\ \ a_{2}=191,\ \ a_{3}=203,\ \
a_{4}=210,\ \ a_{5}=216\ (Sept. \; {\sl 1929}). 
$$

From condition (1), the polinomial coefficients, for $i=0,1,...,5$, are 
$$
A_{1}=26.683,\quad A_{2}=-49.208,\quad A_{3}=28.333,\quad A_{4}=-6.292,\quad
A_{5}=0.483. 
$$
Two higher-order self-similar exponential approximants write  
$$
f_{4}^{*}(t,\tau )=a_{0}\exp \left( \frac{A_{1}}{a_{0}}t\exp \left( \frac{
A_{2}}{A_{1}}t\exp \left( \frac{A_{3}}{A_{2}}t\exp \left( \frac{A_{4}}{A_{3}}
\tau t\right) \right) \right) \right) , 
$$
$$
f_{5}^{*}(t,\tau )=a_{0}\exp \left( \frac{A_{1}}{a_{0}}t\exp \left( \frac{%
A_{2}}{A_{1}}t\exp \left( \frac{A_{3}}{A_{2}}t\exp \left( \frac{A_{4}}{A_{3}}%
t\exp \left( \frac{A_{5}}{A_{4}}\tau t\right) \right) \right) \right)
\right) .  
$$
The estimates for the index at $t=6$ are $f_{4}^{*}(6,1)=194.884$ and  
$f_{5}^{*}(6,1)=206.722.$ The corresponding mapping multipliers, 
$M_{4}(6)=-0.025$ and $M_{5}(6)=0.021$, signal that the subsequence 
$f_{4}^{*},\; f_{5}^{*}$ is stable. From the minimal difference 
condition $\min_{\tau }\left| f_{5}^{*}(6,\tau)-f_{4}^{*}(6,\tau )\right| $ 
one has 
$$
\tau =\exp \left( \frac{6A_{5}}{A_{4}}\tau \right) .  
$$
This gives $\tau =0.7182$. The self-similar exponential approximation 
is $f_{5}^{*}(6)=201.692$. The actual value of the index in {\it October 
1929} was $194$. The percentage error equals $3.965\%$. It is worth 
noting that our lower bound estimate, $f_{4}^{*}(6,1)$, practically 
coincides with the real value.

$(ii)$ Consider the behavior of the industrial share prices in the New York
Stock Exchange characterized by the average Standard\&Poor index, in the
period of time from the {\sl second quarter} of ${\sl 1986}$ to the 
{\sl third quarter} of ${\sl 1987}$ ({\sl 1980}=100):
$$
a_{0}=199.4\ ({\sl II,86}),\ ~a_{1}=198.2,\ ~a_{2}=201.3,\ ~a_{3}=235.1,\
~a_{4}=253,~\ a_{5}=277.5\ ({\sl III,87}). 
$$
And let us make a forecast for the {\it fourth quarter of 1987}. This case
can be considered by analogy with $(i)$. The coefficients of the polynomial
are 
$$
A_{1}=52.12,\quad A_{2}=-103.717,\quad A_{3}=64.096,\quad
A_{4}=-14.883,\quad A_{5}=1.184. 
$$
For the index we obtain the values $f_{4}^{*}(6,1)=202.109$ and 
$f_{5}^{*}(6,1)=226.696$. The mapping multipliers are $M_{4}(6)=-0.01$  
and $M_{5}(6)=0.012$, so we conclude that the sequence of approximants 
locally converges. From the minimal difference condition, we find 
$\tau =0.7119$. The self--similar exponential approximation becomes 
$f_{5}^{*}(6)=215.293$. The actual value of the index in the fourth 
quarter of $1987$ was $218.3$. The forecast error is $1.377\%$.

$(iii)$ Consider the behavior of the Standard\&Poor index in the period of
time from {\sl February} ${\sl 1990}$ to {\sl July} ${\sl 1990}\ ({\sl 
1985}=100)$,
$$
a_{0}=183.4\ 
(Feb.\; {\sl 90}),~a_{1}=188.5,~a_{2}=189.2,~a_{3}=196.4,~a_{4}=202.8,
~a_{5}=204.9\ (Jul.\; {\sl 90}). 
$$
Let us make a forecast for {\sl August 1990}, the time of the Persian Gulf
crisis. The coefficients of the polynomial can be restored giving 
$$
A_{1}=19.883,\quad A_{2}=-25.158,\quad A_{3}=12.783,\quad A_{4}=-2.592,\quad
A_{5}=0.183. 
$$
Repeating the same procedure as above, we find the exponential 
approximants $f_{4}^{*}(6,1)=188.986$ and $f_{5}^{*}(6,1)=201.789$ and 
the mapping multipliers $M_{4}(6)=-0.066$ and $M_{5}(6)=0.043$. From 
the minimal difference condition, we obtain $\tau =0.7327$. The
self-similar exponential approximation is $f_{5}^{*}(6)=197.476$. The actual
value of the index in {\sl August} {\sl 1990} was $188.1$. The percentage
error of our forecast equals\ $4.985\%$, and the lower bound given by 
$f_{4}^{*}(6,1)$ practically coincides with the actual value.

\section{Attempt of Prediction}

Finally, consider the behavior of the Standard\&Poor, Dow Jones, and NYSE
Composite indices in the period of time from {\sl April 30}, ${\sl 1997}$ 
to {\sl September 30}, ${\sl 1997}$, and let us make a forecast for {\sl
October 31, 1997}. 

For the {\bf NYSE Composite} index we have (from the NYSE Historical 
Statistics Archive in Internet): 
$$
a_{0}=416.94\ (Apr.\; 30),\quad a_{1}=441.78,\quad a_{2}=462.44,
$$
$$
a_{3}=494.5,\quad a_{4}=470.48,\quad a_{5}=497.23\ (Sept.\; 30).
$$
The coefficients of the polynomial are 
$$
\quad A_{1}=104.366,\quad A_{2}=-155.195,\quad A_{3}=98.434,\quad
A_{4}=-24.91,\quad A_{5}=2.145.
$$
Repeating the same procedure as above, we find the exponential
approximants $f_{4}^{*}(6,1)=430.124$ and $f_{5}^{*}(6,1)=520.109$ and 
the mapping multipliers $\ M_{4}(6)=-0.022$ and $M_{5}(6)=0.03$. From 
the minimal difference condition we obtain $\tau =0.6974$. The 
self-similar exponential approximation is $f_{5}^{*}(6)=478.855$.

For the {\bf Standard\&Poor} index (data are taken from the DBC Online 
chart) we have 
$$
a_{0}=783\ (Apr.\; 30),~a_{1}=845,~a_{2}=888,~a_{3}=948,~a_{4}=900,~a_{5}=950\
(Sept.\; 30). 
$$
The coefficients of the polynomial are 
$$
A_{1}=222.15,\quad A_{2}=-306.292,\quad A_{3}=189.75,\quad
A_{4}=-47.708,\quad A_{5}=4.1. 
$$
We find the higher exponential approximants $f_{4}^{*}(6,1)=818.966$ 
and  $f_{5}^{*}(6,1)=1030$ and for the mapping multipliers we obtain 
$M_{4}(6)=-0.021$ and $M_{5}(6)=0.028$. From the minimal difference 
condition, we find $\tau =0.6978$. Thus, the self--similar exponential  
approximation is $f_{5}^{*}(6)=935.082$.

Similarly, for the {\bf Dow Jones} index (data are taken from the DBC Online 
chart) one has
$$
a_{0}=6950\
(Apr.\; 30),~a_{1}=7300,~a_{2}=7700,~a_{3}=8200,~a_{4}=7670,~a_{5}=8000\
(Sept. \; 30), 
$$
which gives the polynomial coefficients 
$$
A_{1}=1477,\quad A_{2}=-2291,\quad A_{3}=1528,\quad A_{4}=-399.167,\quad
A_{5}=35. 
$$
Following the same steps as above, we find the exponential approximants 
$f_{4}^{*}(6,1)=7108$ and $f_{5}^{*}(6,1)=8400$ and the mapping 
multipliers $\ M_{4}(6)=-0.025\ $and $M_{5}(6)=0.04$. From the minimal
difference condition we obtain $\tau =0.694088$. The self--similar
exponential approximation is $f_{5}^{*}(6)=7788$.

In conclusion, we have shown that resummation techniques of theoretical
physics can be successfully employed for analyzing time series. The most
convenient such a technique, to our mind, is the algebraic self-similar
renormalization [15-17]. We have applied this approach to analyzing time
series corresponding to different stock markets. From our point of view,
market crises are somewhat similar to critical phenomena in statistical 
physics. There exists a temporal region around a crisis, where the
behaviour of a market, being intimately connected with its behaviour at 
times relatively distant from visible anomalies [19], begins exhibiting 
specific features, like log--periodic oscillations [3-5]. Such a precrisis
region around a phase transition, and precrisis effects are similar to 
precursor phenonema caused by heterophase fluctuations [20]. The point of 
view that market crises have their origin in the collective behaviour of 
many interacting agents and that the stock market crises are analogous to 
critical phenomena has also been promoted by other researchers [5]. The
similarity between the crises of stock markets and the critical phenomena
of statistical systems makes it possible to apply for their description
the same resummation methods, such as the algebraic self--similar 
renormalization [15-17]. Farther development of this approach will be 
presented in the following papers.

\end{document}